\documentclass{article}
\usepackage{spconfa4,amsmath,amssymb,graphicx}
\usepackage{caption}
\usepackage{subcaption}
\usepackage{xcolor}
\usepackage{booktabs}    
\usepackage{adjustbox}   
\usepackage{enumitem}
\usepackage{bm}
\usepackage{multirow}
\usepackage{makecell}
\usepackage{hyperref}
\usepackage{xurl}
\newif\iftodo
\todotrue

\title{Singer-Informed Vocal Source Separation\\for Multi-Singer Music Mixtures}
\name{Jocelyn Xu and Minje Kim\thanks{This material is based upon work supported by the National Science Foundation under Grant No. 2512987.} } 

\address{University of Illinois Urbana-Champaign, Siebel School of Computing and Data Science}
\begin{document}
\ninept
\maketitle
\begin{abstract}
Music source separation systems typically extract a single vocal track and do not distinguish between multiple singers. We study singer-informed vocal source separation for multi-singer mixtures. Our framework introduces a short enrollment recording of a target singer to guide separation through a learned embedding. The singer embedding is incorporated using feature concatenation or feature-wise linear modulation (FiLM), enabling the model to focus on the target singer while suppressing interference. We construct a duet dataset based on DAMP-VSEP with quality filtering and non-overlapping enrollment segments. Experiments on solo and duet settings show that while baseline models perform well for single-singer mixtures, the proposed method improves target-singer extraction in multi-singer cases, increasing target-singer SI-SDR from $0.33$ dB to $5.58$ dB. Fréchet Audio Distance (FAD) further shows improved perceptual quality and better alignment with target audio distributions. Code and checkpoints are available at \url{https://github.com/jocelynxu01/singer-separation-paper}.


\end{abstract}

\section{Introduction}
\label{ch:intro}


Music source separation aims to decompose a mixed audio recording into its individual sources, such as vocals and instrumental components. Among these tasks, vocal source separation has received significant attention due to applications including music remixing, karaoke generation, and audio editing \cite{stoter2019open, defossez2019demucs, defossez2021hybrid, hennequin2020spleeter}.

Recent deep learning approaches, such as Open-Unmix \cite{stoter2019open}, have demonstrated effective vocal extraction by learning spectrogram masks from mixture signals. These systems typically assume that the mixture contains a single vocal source together with accompaniment, although, in real-world recordings, multiple singers often perform simultaneously. In duet or ensemble scenarios, vocal sources overlap in time and frequency, making it difficult for conventional models to isolate individual singers. As a result, standard vocal separation systems often reconstruct the combined vocal signal rather than distinguishing between different singers.

One promising strategy is to incorporate identity information about the target singer. In speech processing, it is common for a \textit{target speaker extraction} (TSE) system to use a short enrollment recording to guide separation \cite{wang2019voicefiltertargetedvoiceseparation}. By conditioning the model on the target voice, the system can focus on extracting the desired speaker while suppressing interfering voices, as well as other non-speech interferences. Extending this idea to singing voice separation leads to a vocal separation result that suppresses or removes only the target singer, while the other singers and background music are all considered interferences. However, this configuration is challenging due to the limited availability of datasets containing isolated singing voices and multi-singer mixtures, while clean singer-specific enrollment signals are hard to acquire in most real-world cases. In addition, singing voice signals tend to contain a greater variation in pitch, timbre, and vocal expression, making the conditioning process challenging. These challenges motivate the development of singer-informed separation methods that incorporate singer identity information for multi-singer vocal separation.


To address this problem, a singer-informed separation framework is introduced. We use a short enrollment recording of the target singer to extract a singer embedding that conditions the separation model during inference. By learning singer embeddings directly from enrollment audio, the conditioning mechanism captures singer-specific characteristics despite variations in pitch and vocal expression, allowing the separation model to focus on the target singer. We ensure that the system is robust to background music present in the enrollment signal, so that the conditioning mechanism is effective once the enrollment signal contains only the target singer. In the duet cases, the singer embedding is particularly helpful in distinguishing between multiple singers present in the mixture.







The main contributions of this work are:
\begin{itemize}[leftmargin=*, itemindent=0pt, itemsep=0pt, topsep=0pt] 
\item A singer-informed vocal separation framework that leverages enrollment audio to extract a target singer from multi-singer mixtures (up to two singers in this study).
\item A singer embedding model for learning singer identity representations from short vocal segments.
\item A duet dataset construction pipeline based on DAMP-VSEP \cite{smule2019damp} with quality filtering and non-overlapping segment selection.
\item Experimental evaluation on solo and duet mixtures, including analysis of different conditioning strategies.

\end{itemize}

\section{Related Work}
\label{ch:related}

Music source separation (MSS) aims to decompose a mixed audio signal into its constituent sources, such as vocals and instrumental components. Deep learning methods have significantly advanced MSS. Open-Unmix predicts time-frequency masks from magnitude spectrograms \cite{stoter2019open}, while Demucs operates directly on raw waveforms using an encoder--decoder architecture \cite{defossez2019demucs}. Hybrid approaches that combine waveform and spectrogram representations have further improved performance \cite{defossez2021hybrid}. More recent MSS methods, such as Moises-Light \cite{hung2025moises} and Task-Aware Unified Source Separation \cite{Saijo2025_tuss}, have continued to advance separation quality and efficiency, while singer-informed vocal separation was not addressed. Most systems separate predefined stems and treat all vocal content as a single source.

Vocal source separation focuses on isolating singing voices from accompaniment, but most systems extract a single vocal stem rather than separating individual singers \cite{stoter2019open,defossez2019demucs}. Recent work has begun to address multiple singing voices separation (MSVS), including karaoke-oriented systems \cite{lin2024singer}, benchmark datasets such as MedleyVox and IdolSongsJP \cite{jeon2023medleyvox, Suda2025IdolSongsJP}, and models addressing data scarcity and inter-singer correlation \cite{jung2026unmixx}. However, these approaches remain largely blind and do not support extraction of a specific target singer. Inconsistency in identity over time has also been observed in separated outputs \cite{yu2023zero}, motivating methods that incorporate explicit identity information.

Speaker-informed speech separation extracts a target speaker using enrollment audio. Methods such as SpeakerBeam \cite{vzmolikova2019speakerbeam}, VoiceFilter \cite{wang2019voicefiltertargetedvoiceseparation}, and SpEx+ \cite{ge2020spex+} incorporate speaker embeddings to guide separation, as well as more recent generative approaches \cite{navon2025flowtse,hsieh2025adaptivedeterministicflowmatching}. While effective in speech, applying these techniques to singing remains challenging due to greater variability and limited labeled data.

Many approaches use embedding-based conditioning to incorporate identity information. Embeddings are typically learned via contrastive objectives \cite{bromley1993signature} or derived from internal representations \cite{ge2020spex+}. These embeddings can be incorporated through feature concatenation or more flexible methods such as feature-wise linear modulation (FiLM) \cite{perez2018film}, enabling models to focus on the target singer.

Overall, prior work on MSS and vocal separation treats vocals as a single source, and speaker-informed methods are studied in speech. This work bridges these areas by introducing singer-informed framework for target singer extraction in multi-singer mixtures.

\section{Dataset Construction}
\label{ch:dataset}

\subsection{Source Dataset: DAMP-VSEP}

We construct our dataset based on DAMP-VSEP \cite{smule2019damp}, which contains user-generated singing recordings with backing tracks. The dataset captures realistic variation in vocal characteristics and recording conditions, making it suitable for singing voice separation. However, DAMP-VSEP is not directly suitable for singer-conditioned duet separation due to the limited number of duet recordings with clean vocals and valid enrollment segments. In addition, DAMP-VSEP does not provide singer identities across songs, each song is treated as a unique singer. To address this, we develop a dataset construction pipeline that filters recordings, extracts vocal segments, and generates additional duet mixtures.

The original dataset contains amateur singers from 155 countries, 36 languages, 6,456 artists, and 11,494 unique compositions, with a total of 41,768 recordings. After basic duplicate cleaning, 41,749 unique recordings are retained, including 20,849 solo and 20,900 duet performances.

\subsection{Data Filtering and Enrollment Part Extraction}

Due to the user-created nature of the DAMP-VSEP dataset, the ground-truth singing voice tracks sometimes contain background music when the user recorded singing while playing the backing track on a loudspeaker. To improve data quality, we apply the non-intrusive speech-quality estimator, DNSMOS \cite{reddy2021dnsmos}, to exclude noisy or distorted vocal tracks. We use a threshold of 3.0 to identify clean vocal recordings. Duet recordings are kept only if both singers meet the threshold. 
After applying DNSMOS filtering, 14,381 solo recordings and 6,389 duet recordings are retained. This yields a cleaner dataset for later processing.


After filtering, voice activity detection (VAD) \cite{ramirez2004efficient} is applied to extract candidate singing segments. An energy-based method identifies vocal frames using a normalized energy threshold of $0.001$. The vocal frames are then grouped into segments with gaps of up to $0.05$ seconds tolerated during initial detection and neighboring segments within $0.15$ seconds merged to reduce fragmentation. Only segments longer than 3 seconds are retained as enrollment candidates. After this step, 13,194 solo performances and 5,578 duet songs contain at least one valid enrollment segment. The rest of the singing voice signals are used to form training and testing mixtures: we ensure that the mixture audio is not contaminated by the enrollment portions.

\subsection{Duet Mixture Generation}

Since valid duet recordings are significantly limited after pre-processing steps, we generate additional synthetic duet mixtures by pairing vocal recordings from different singers. Each vocal track is treated as an anchor and additional partner singers are randomly sampled to increase diversity.

For each anchor singer, we generate additional duet mixtures by pairing the anchor vocal with 10 different partner vocal recordings while using the same background track. Before mixing, each vocal source is scaled to a consistent loudness level ($-20$ dB), and the background track is slightly reduced in volume to avoid overpowering the vocals ($-26$ dB). The signals are then aligned in length, summed to form the mixture. This process significantly increases the number of duet training examples to 55,780.

The dataset is divided into training, validation, and test sets using a fixed random seed with an 8:1:1 ratio. However, for the duet test set, we eliminated random mixtures and used only the original duet recordings (559 songs) to measure the proposed method's performance on the real-world data.

\section{Methodology}
\label{ch:method}

\subsection{Problem Formulation}

Given a mixture signal $\bm x \in\mathbb{R}^N$ of length $N$ containing vocal and background components, the goal is to estimate the target vocal signal. In the general case, the mixture may contain multiple singers:
$\bm x = \sum_{k=1}^K \bm s^{(k)} + \bm b$,    
where $\bm s^{(k)}$ denotes the vocal signal of the $k$-th singer and $\bm b$ is the background accompaniment. In this formulation, each $\bm s^{(k)}$ corresponds to a single singer, even if the vocal signal as a sum is polyphonic. In this work, $K$ is limited to 2.

In multi-singer scenarios, a short enrollment signal $\bm e^{(\kappa)}$ from the target singer $\kappa=\{1,\ldots,K\}$ is provided. A singer embedding function $f(\cdot)$ extracts a fixed-dimensional representation $\bm z^{(\kappa)} = f(\bm e^{(\kappa)})$, which is used to condition the separation model:
\begin{equation}\label{eq:sep_model}
\hat{\bm s}^{(\kappa)} = g(\bm x, \bm z^{(\kappa)}),    
\end{equation}
where $g(\cdot)$ denotes the singer-informed separation model. In the single-singer case, the conditioning information $\bm z^{(\kappa)}$ does not necessarily provide any additional information, thus reducing eq. \eqref{eq:sep_model} to $\hat{\bm s} = g(\bm x)$ if necessary, by having $\bm s$ as the only singer. 

\subsection{Framework Overview}

\begin{figure}[t]
    \centering
    \includegraphics[width=\columnwidth]{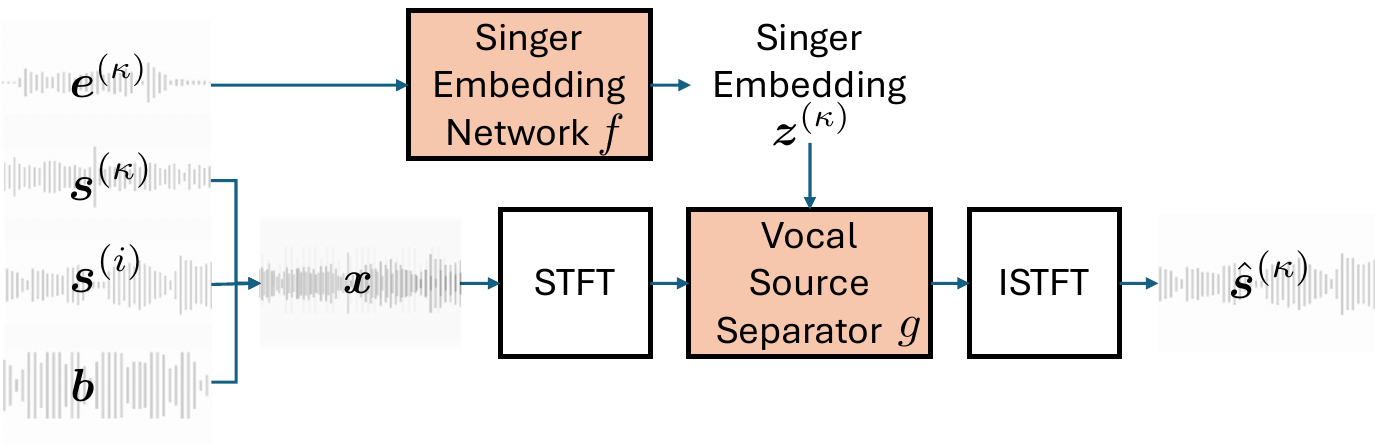}
    \vspace{-0.2in}
    \caption{Overview of the singer-informed vocal separation model.}
    \vspace{-0.1in}
    \label{fig:framework}
\end{figure}

The proposed system combines a singer embedding model with a neural vocal separation model, as illustrated in Fig.~\ref{fig:framework}. The mixture waveform is first transformed into a time-frequency representation, and a separation network predicts the target vocal component.

In multi-singer cases, an enrollment segment from the target singer is passed through the embedding network to produce the singer representation. This embedding is used to condition the separation model, allowing it to focus on the target singer $\kappa$ while suppressing interfering vocals. When $K=2$, we denote the other singer by $i$, i.e., $\bm x = \bm s^{(\kappa)} + \bm s^{(i)}+ b$.

\subsection{Singer Embedding}

A singer embedding model is learned to extract identity information from short enrollment audio. The design of the embedding model is inspired by the speaker representation learning approach proposed in  \cite{sivaraman2021zeroshotpersonalizedspeechenhancement}. Each input segment is converted to a magnitude spectrogram and processed by a GRU network. The final hidden state is used as a fixed-dimensional embedding representing the target singer. Contrastive learning forces the embeddings of the same singer to be similar, while those of different singers differ. These embeddings are used to condition the separation model in multi-singer scenarios.

\subsection{Separation Model}

We build upon the Open-Unmix architecture \cite{stoter2019open}, which operates on magnitude spectrograms and predicts a spectral mask for source separation. The estimated magnitude is combined with the mixture phase to reconstruct the waveform. While effective for single-singer separation, this model does not distinguish between multiple vocal sources. Our proposed singer conditioning addresses this limitation.

\subsection{Speaker Conditioning}

\begin{figure}[t]
    \centering
    \vspace{-0.1in}
    \includegraphics[width=\columnwidth]{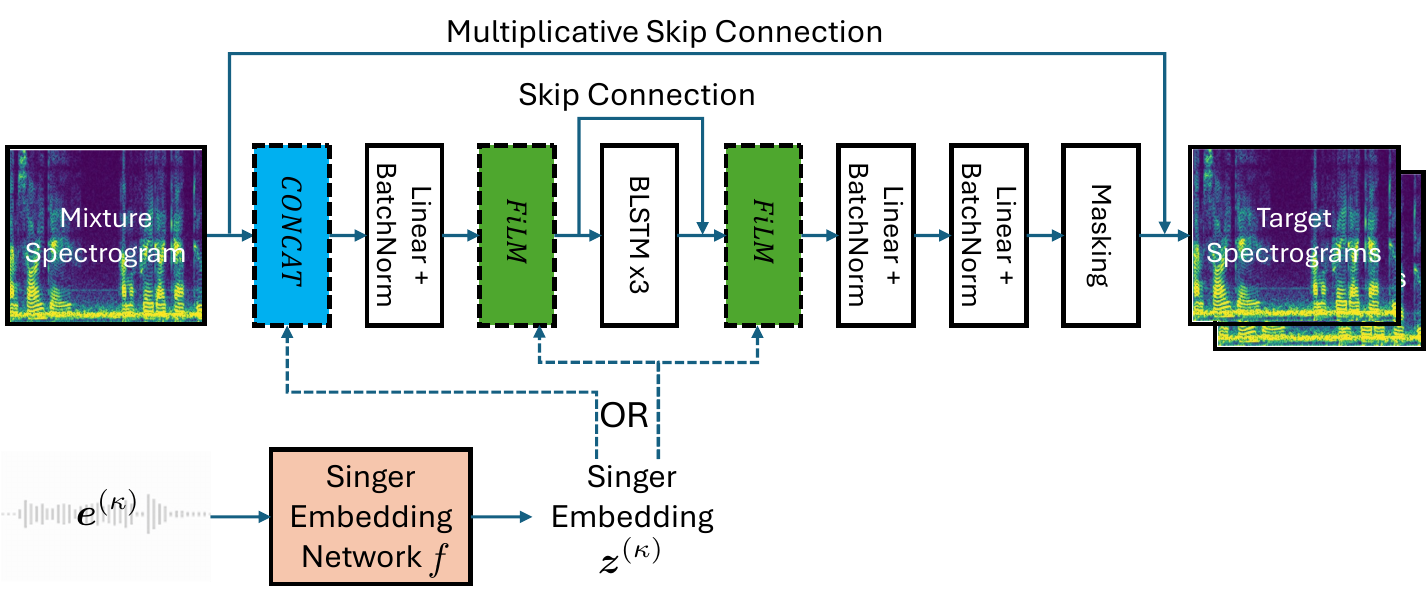}
    \caption{The singer embedding is incorporated through one of two conditioning strategies: concatenation at the input feature or FiLM-based modulation at intermediate layers. Blue box is used in the concatenation setting, and green boxes are used in the FiLM setting. }
    \vspace{-0.1in}
    \label{fig:embedding_details}
\end{figure}

We explore two approaches to incorporate singer embeddings into the separation model, illustrated in Fig.~\ref{fig:embedding_details}.

\noindent \textbf{Concatenation.}
The singer embedding is repeated across time and concatenated with input features before the separation network. This allows models to incorporate singer identity information at input level.

\noindent \textbf{FiLM Conditioning.}
We also apply feature-wise linear modulation (FiLM) \cite{perez2018film} to modulate intermediate feature representations based on the target singer information:
\[
\text{FiLM}\big(\bm h, \bm z^{(\kappa)}\big) = \gamma\big(\bm z^{(\kappa)}\big) \odot \bm h + \beta\big(\bm z^{(\kappa)}\big),
\]
where $\bm h$ is a feature representation in the separation module and $\gamma\big(\bm z^{(\kappa)}\big)$ and $\beta\big(\bm z^{(\kappa)}\big)$ are learned functions of the singer embedding. Compared to concatenation, FiLM provides a more flexible mechanism for conditioning, allowing speaker information to influence both low-level and high-level features.

\subsection{Training Objective}
The separation model is trained using the scale-invariant signal-to-distortion ratio (SI-SDR) loss \cite{le2019sdr}. Given the ground truth vocal $\bm s^{(\kappa)}$ and estimate $\hat {\bm s}^{(\kappa)}$, the model is optimized by minimizing the negative SI-SDR. In multi-singer settings, the target corresponds to the selected singer, while all other components, i.e., $\bm b + \sum_{k \in \{1,\ldots,K\}\setminus\{\kappa\}} \bm s^{(k)}$, are treated as interference.  We also investigate a dual loss variant that additionally supervises the residual signal. Details are presented in Section~\ref{ch:results}.

\section{Experimental Setup}
\label{ch:experiments}

\subsection{Training Setup}

\textbf{The Singer Embedding Model} consists of a two-layer GRU model (hidden size 32) operating on magnitude spectrograms. The final hidden state is used as the embedding, and similarity is computed via inner product. The model is trained using binary cross-entropy loss by having samples from the same singer as the positive class and those from different singers are treated as the negative class.

\noindent\textbf{The Vocal Separation Model} is based on Open-Unmix. Input audio is converted to mono and processed using STFT. Training is performed on 6-second segments using Adam with learning rate $10^{-3}$ and weight decay $10^{-5}$. Learning rate decay and early stopping are applied based on validation performance.

\noindent\textbf{Singer Conditioning} uses one of two methods. First, embeddings are repeated across time and concatenated with the input spectra. Second, the FiLM module estimates scaling and bias parameters from the enrollment to modulate the separator's features.


\subsection{Evaluation Setup}

We evaluate the proposed speaker-informed separation framework under both solo and duet mixtures. 

\noindent\textbf{Solo Performances:} The predicted vocal is evaluated against the ground-truth vocal, and the residual against the background track. This measures vocal reconstruction and background suppression.

\noindent\textbf{Duet Performances:} In the duet setting, the model is conditioned on the target singer and is expected to extract only the target vocal. Hence, the residual is evaluated against the sum of the interfering singer and background music. Additional comparisons (e.g., against the interfering singer) provide a diagnostic analysis.

\begin{table*}[t]
\centering
%
\begin{minipage}[t]{0.58\textwidth}
\begin{adjustbox}{width=.95\columnwidth}
\begin{tabular}{lllrrrr}
\toprule
\multirow{2}{*}{System} & \multirow{2}{*}{Loss} & \multirow{2}{*}{Conditioning} 
& \multicolumn{4}{c}{Source in Comparison (dB $\uparrow$)} \\
\cmidrule(lr){4-7}
& & & $\bm s^{(\kappa)}$ & $\bm s^{(\kappa)}+\bm s^{(i)}$ & $\bm b$ & $\bm s^{(i)} +\bm b$ \\

\midrule
\multirow{7}{*}{\makecell{Proposed\\models}} & $\mathcal{L}_\text{v}$ & FiLM

& $3.39$ & $0.17$ & $-10.14$ & $-2.99$ \\

& $\mathcal{L}_\text{v}$ & Concat.
& $2.92$ & $-1.61$ & $-6.85$ & $5.04$ \\

\cmidrule(lr){2-7}

& $\mathcal{L}_\text{v+r}$, $\lambda=0.05$ & FiLM
& $3.15$ & $-0.39$ & $-6.55$ & $6.08$ \\

& $\mathcal{L}_\text{v+r}$,  $\lambda=0.1$ & FiLM
& $1.85$ & $1.39$ & $-6.45$ & $4.26$ \\

& $\mathcal{L}_\text{v+r}$, $\lambda=0.2$ & FiLM
& $2.55$ & $1.57$ & $-5.94$ & $4.85$ \\

& $\mathcal{L}_\text{v+r}$, $\lambda=0.05$ & Concat.
& $1.92$ & $-2.58$ & $-6.78$ & $6.96$ \\

& $\mathcal{L}_\text{v+r}$, $\lambda=0.1$ & Concat.
& $\mathbf{5.58}$ & $0.59$ & $-5.91$ & $\mathbf{7.98}$ \\

\midrule

Baseline
& $\mathcal{L}_\text{v}$
& -
& $0.33$ & $15.88$ & $-1.71$ & $1.47$ \\

\bottomrule
\end{tabular}
\end{adjustbox}
\caption{Results on the duet test setting (SI-SDR).}
\label{tab:duet_results}
\end{minipage}\hspace{-0.1in}
\begin{minipage}[t]{0.2\textwidth}
\begin{adjustbox}{width=.975\columnwidth}
\begin{tabular}{rr}
\toprule
\multicolumn{2}{c}{Source in Comparison (dB $\uparrow$)} \\
\cmidrule(lr){1-2}
$\bm s$ & $\bm b$ \\
\midrule

 $7.26$ & $-8.36$ \\

 $1.87$ & $-0.34$ \\

\cmidrule(lr){1-2}

 $6.34$ & $2.61$ \\

 $5.16$ & $0.60$ \\

 $7.48$ & $2.19$ \\

 $-1.75$ & $0.81$ \\

 $9.26$ & $5.31$ \\

\midrule

 $17.80$ & $16.68$ \\

\bottomrule
\end{tabular}
\end{adjustbox}
\caption{Results on the solo test setting (SI-SDR).}
\label{tab:solo_results}

\end{minipage}\hfill
\begin{minipage}[t]{0.21\textwidth}

\centering
\begin{adjustbox}{width=.975\columnwidth}
\begin{tabular}{rr}
\toprule
 \multicolumn{2}{c}{Source in Comparison (FAD $\downarrow$)} \\
\cmidrule(lr){1-2}
$\bm s^{(\kappa)}$ & $\bm s^{(i)} + \bm b $ \\
\midrule

 $7.15$ & $1.47$ \\

 $\mathbf{0.37}$ & $0.18$ \\

\cmidrule(lr){1-2}

 $1.07$ & $0.25$ \\

 ${0.63}$ & $0.10$ \\		

 $0.82$ & $0.17$ \\

 $1.47$ & $0.13$ \\

 ${0.46}$ & $\mathbf{0.06}$ \\

\midrule

 $2.28$ & $2.44$ \\

\bottomrule
\end{tabular}
\end{adjustbox}

\caption{Results on the duet test setting (EnCodec FAD).}
\label{tab:duet_results_fad_encodec}
\end{minipage}

\end{table*}

\subsection{Evaluation Metrics}

We evaluate separation performance using both reconstruction-based and perceptual metrics:

    \noindent \textbf{Scale-Invariant Signal-to-Distortion Ratio (SI-SDR)} \cite{le2019sdr}: Measures similarity between estimated and reference signals while being invariant to global scaling. Used for both training and validation.  
    
    \noindent \textbf{Frechet Audio Distance (FAD)} \cite{kilgour2019frechetaudiodistancemetric}: A perceptual metric that measures the distance between generated and reference audio distributions in a learned embedding space. For testing, it complements SI-SDR by capturing perceptual quality beyond sample-level reconstruction, especially since our target signals are noisy.

\section{Results and Discussion}
\label{ch:results}


Tables~\ref{tab:duet_results}, \ref{tab:solo_results}, and \ref{tab:duet_results_fad_encodec} summarize results for baseline Open-Unmix and singer-conditioned models using concatenation or FiLM. All proposed models are trained on the constructed duet dataset described in Section~\ref{ch:dataset}. We use the original Open-Unmix model following its standard training configuration as the baseline. Since it does not use singer conditioning, it separates the vocal stem rather than a specific target singer. During evaluation, only original duet recordings are used to better reflect realistic and challenging singing scenarios. The vocal loss $\mathcal{L}_\text{v}$ refers to $-\text{SI-SDR}(\bm s^{(\kappa)} \,\|\, \hat {\bm s}^{(\kappa)})$, while the dual loss $\mathcal{L}_\text{v+r}$ additionally compares the reconstruction quality of the residual: $-\text{SI-SDR}(\bm s^{(\kappa)} \,\|\, \hat {\bm s}^{(\kappa)}) + \, -\lambda \text{SI-SDR}(\bm s^{(i)} + \bm b \,\|\, \bm x - \hat {\bm s}^{(\kappa)})$, where $\bm s^{(i)}$ denotes the interfering singer. In the duet setting, the additional term in the residual could potentially improve the separation quality. The resulting numbers are on the test fold, where we report SI-SDR numbers by flipping the sign of the loss functions, i.e., higher values indicate better separation.

\subsection{SI-SDR Results}

In the duet setting (Table \ref{tab:duet_results}), the baseline behaves like a generic vocal separator. It performs poorly ($0.33$ dB) compared to the target singer ($\bm s^{(k)}$), but strongly ($15.88$ dB) in the combined vocal signal ($\bm s^{(\kappa)}+\bm s^{(i)}$), indicating that it extracts both singers together rather than isolating the target.

In contrast, singer-conditioned models achieve better results in target-singer separation. FiLM with target vocal-only loss improves the target vocal separation to $3.39$ dB, while concatenation-based conditioning achieves $2.92$ dB, both significantly better than the baseline ($0.33$ dB). At the same time, their comparison against the sum of both singers, $\bm s^{(\kappa)}+\bm s^{(i)}$, scores remain weaker (e.g., $0.17$ dB for FiLM and $-1.61$ dB for concatenation), confirming that they no longer reconstruct the non-target vocal as expected.

Residual evaluation further highlights this behavior. The baseline's residual effectively predicts the background music $\bm b$, hence it achieves only $1.47$ dB compared to $\bm s^{(2)} +\bm b$, while conditioned models reach much higher values, with the best result of $7.98$ dB obtained by concatenation with dual loss at $\lambda=0.1$. Other strong settings include concatenation at $\lambda=0.05$ ($6.96$ dB) and FiLM at $\lambda=0.05$ ($6.08$ dB). Meanwhile, the baseline's performance against the right target residual, $\bm b$, is also poor, as we have not used the loss on the residual reconstruction. The poor performance of the proposed models on reconstructing $\bm b$ is also expected, as they have not been trained to only reconstruct the background music.

Both FiLM and concatenation improve duet settings over the baseline, with concatenation achieving a stronger residual reconstruction while FiLM being more consistent.

In the solo setting (Table \ref{tab:solo_results}), the baseline outperforms our proposed singer-conditioned models, achieving $17.80$ dB for the solo vocal reconstruction and $16.68$ dB for background. This is expected since singer conditioning is unnecessary for solo singing.

\subsection{FAD Results}














Due to the originally noisy nature of the dataset (e.g., predicted sources can sometimes be cleaner than the ground truth target), we found that SI-SDR values can be worse than the perceptual quality of the separation. Hence, we report FAD scores in Table~\ref{tab:duet_results_fad_encodec}, computed using EnCodec \cite{DefossezA2023encodec} embeddings. Lower values indicate closer alignment between predicted and reference audio distributions. The FAD computed using mixtures as predictions is $28.38$.

The baseline model produces relatively high vocal and residual FAD scores, $2.28$ and $2.44$ respectively, indicating that the predicted vocal remains far from the sources. This behavior is consistent with the SI-SDR results. In contrast, the singer-conditioned models substantially reduce both vocal and residual FAD scores. For example, concatenation with vocal-only loss achieves a vocal FAD of $0.37$ and a residual FAD of $0.18$. Dual-loss models further improve performance, with the lowest residual FAD achieved by concatenation with $\lambda=0.1$, which reaches $0.06$. Again, the low FAD scores indicate that the target singer is correctly extracted, leaving the interfering singer and background music in the residual as desired. 
Overall, the FAD results support the SI-SDR findings: singer conditioning improves target-singer separation in duet mixtures.

\section{Conclusion}
\label{ch:conclusion}

This paper presented a singer-informed vocal separation framework for extracting a target singer from multi-singer mixtures. By incorporating singer embeddings into the separation model through concatenation and FiLM conditioning, the proposed approach shifted the model from generic vocal separation toward target-singer extraction. Experimental results show that while the baseline performs well in solo settings, the singer-conditioned models achieve more effective target-singer separation in duet mixtures, producing improved reconstruction of the target singer and more accurate residual signals. Future work will extend the method to more singers, noisy enrollment, and more advanced generative architectures.

\bibliographystyle{IEEEbib}
\bibliography{refs}

\end{document}